\documentclass[12pt,epsf,epsfig]{article}
\usepackage{graphicx}
\usepackage{amsmath,latexsym}

\begin{document}

\begin{center}
{\huge{String Model with Mesons and Baryons in Modified Measure Theory.}} \\
\end{center}

\begin{center}
T.O. Vulfs \textsuperscript{1,2,3}, E.I. Guendelman \textsuperscript{1,3,4} \\
\end{center}

\begin{center}
\textsuperscript{1} Department of Physics, Ben-Gurion University of the Negev, Beer-Sheva, Israel \\
\end{center}

\begin{center}
\textsuperscript{2} Fachbereich Physik der Johann Wolfgang Goethe Universit\"{a}t, Campus Riedberg, Frankfurt am Main, Germany \\
\end{center}

\begin{center}
\textsuperscript{3} Frankfurt Institute for Advanced Studies, Giersch Science Center, Campus Riedberg, Frankfurt am Main, Germany \\
\end{center}

\begin{center}
\textsuperscript{4} Bahamas Advanced Study Institute and Conferences, 4A Ocean Heights, Hill View Circle, Stella Maris, Long Island, The Bahamas \\
\end{center}

E-mail: vulfs@post.bgu.ac.il, guendel@bgu.ac.il

\abstract
We consider string meson and string baryon models in the framework of the modified measure theory, the theory that does not use the determinant of the metric to construct the invariant volume element. As the outcome of this theory, the string tension is not placed ad hoc but is derived. When the charges are presented, the tension undergoes alterations. In the string meson model there are one string and two opposite charges at the endpoints. In the string baryon model there are two strings, two pairs of opposite charges at the endpoints and one additional charge at the intersection point, the point where these two strings are connected. The application of the modified measure theory is justified because the Neumann boundary conditions are obtained dynamically at every point where the charge is located and Dirichlet boundary conditions arise naturally at the intersection point. In particular, the Neumann boundary conditions that are obtained at the intersection point differ from that considered before by 't Hooft in [hep-th/0408148] and are stronger, which appears to solve the nonlocality problem that was encountered in the standard measure approach. The solutions of the equations of motion are presented. Assuming that each endpoint is the dynamical massless particle, the Regge trajectory with the slope parameter that depends on three different tensions is obtained.

\section{Introduction}

When considering the action formulation of a theory the standard measure of the integration, $\sqrt{-g}$ (where $g$ is the determinant of the spacetime metric), is usually used. It must be a density under diffeomorphic transformations and therefore, $\sqrt{-g}$ may not be a unique choice. In our paper, the integration measure is constructed out of two scalar fields. When the measure is modified, an additional degree of freedom is artificially added, then the action is varied with respect to the new dynamical field. As a main result, it becomes possible to prescribe the meaning to the constants that were put by hand before. Modified measure theories \cite{i} are widely used in gravity, for example, for solving the cosmological constant problem \cite{j}, the fifth force problem \cite{k}, the unified dark energy - dark matter problem \cite{z}. Such theories are considered within string theory in \cite{l,m} and in particular with the Galileon modified measure in \cite{n,o}. \\

The most significant contribution of the modified measure theory to string theory is that it shows how to derive the tension instead of putting it ad hoc to the action. The aim of this paper is to apply this development to the construction of string meson and string baryon models of a special configuration that arise naturally in the framework of the modified measure theory. \\


String models of hadrons are well studied in the literature \cite{a,b,c,d,e,f,g,h}. While there is only one possible string configuration to represent a meson, that is, a single string with the opposite charges at both endpoints, the baryons have more freedom. Three strings with the charges at each endpoint can be arranged, for example,  in $\Delta$-model, in the Y-shaped model which requires a vertex. A one-string quark-diquark model is also possible as the limit of the Y-configuration. In this paper, we present new string meson and string baryon models. \\

Our string meson model consists of an open string with the opposite charged endpoints. These charges signify the discontinuity of the string tension and therefore, in this case the termination of the string. As opposite to the standard string theory we put the charges at first, then we see that they must be opposite and then Neumann boundary conditions are obtained. We do not put these conditions at the endpoints but we derive them. \\

Our string baryon model is constructed out of two strings with the opposite charged endpoints each. However, one of them has an additional charge. This charge brings the alteration to the tension. But instead of termination, the string changes its tension value and continues. The difference from the previous models, besides the number of strings used, is that the Dirichlet boundary conditions arise naturally at the intersection point. In \cite{h} boundary conditions are enforced by the Lagrange multipliers and differ from ours, see Section 5. Here, both Dirichlet and Neumann boundary conditions come from the measure initially modified. \\

In Section 2 we provide a general information on a string in the modified measure theory. Section 3 is dedicated to the string meson model, and Neumann boundary conditions are derived. In Section 4 we connect two strings, thereby constructing a baryon, and Dirichlet boundary conditions are applied at the intersection point. It is substantial that these conditions are obtained dynamically. In Section 5 we present the approach, the main results and the problem of \cite{h}. The resolution of this problem is given in Section 6. Section 7 is devoted to the solutions of the equations of motion and to the derivation of the energy and angular momentum of our system. In Section 8 we discuss the consequences for the quantum case. Conclusions are given in Section 9.\\

\section{The Modified Measure String}

The standard sigma-model string action is

\begin{equation} \label{1}
S_{sigma-action} = -T\int d\sigma d \tau \frac12 \sqrt{-\gamma} \gamma^{ab}\partial_a X^{\mu} \partial_b X^{\nu} g_{\mu\nu},
\end{equation}

where $T$ is the tension of the string; $\gamma^{ab}$ is the intrinsic metric on the worldsheet, where the indices are $a, b = 0,1$; $\gamma$ is the determinant of $\gamma_{ab}$; $g_{\mu \nu}$ is the metric of the embedding spacetime with $D$ dimensions, where the indices are $\mu, \nu = 0, \ldots, D$. $X^{\mu}$ are coordinate functions; $X^{\mu} = X^{\mu}(\sigma, \tau)$, where $\sigma, \tau$ are worldsheet parameters. \\

The equations of motion with respect to the dynamical variables $\gamma^{ab}$ and $X^{\mu}$ are

\begin{equation} \label{eq:16}
T_{ab} = (\partial_a X^{\mu} \partial_b X^{\nu} - \frac12 \gamma_{ab}\gamma^{cd}\partial_cX^{\mu}\partial_dX^{\nu}) g_{\mu\nu}=0,
\end{equation}

\begin{equation} \label{eq:17}
\frac{1}{\sqrt{-\gamma}}\partial_a(\sqrt{-\gamma} \gamma^{ab}\partial_b X^{\mu}) + \gamma^{ab} \partial_a X^{\nu} \partial_b X^{\lambda}\Gamma^{\mu}_{\nu\lambda}=0,
\end{equation}

where $\Gamma^{\mu}_{\nu\lambda}$ is the affine connection for the external metric. \\

The only restriction to the integration measure is that it must be a density under diffeomorphic transformations. In this paper $\Phi$ is chosen to be a measure density instead of the standard one $\sqrt{-\gamma}$. By the definition $\Phi$ is

\begin{equation}
\Phi = \frac12 \epsilon_{ij} \epsilon^{ab} \partial_a \varphi^i \partial_b \varphi^j,
\end{equation}

where $\varphi^i (i = 1, 2)$ are two (by the number of dimensions) additional worldsheet scalar fields, $\epsilon_{ij}$ and $\epsilon^{ab}$ are the Levi-Civita symbols. \\

Then

\begin{equation}
S = \int d\sigma d \tau \Phi L,
\end{equation}

where $L$ is an arbitrary Lagrangian, transforming as a scalar under general coordinate transformations. \\

The variation with respect to $\varphi_i$ is

\begin{equation}
\epsilon^{ab} \partial_b \varphi_j \partial_a L = 0.
\end{equation}

Since $\det(\epsilon^{ab}\partial_b \varphi_j) \sim \Phi$, then, if $\Phi \ne 0$, it leads to the condition

\begin{equation} \label{eq:50}
L = M = const.
\end{equation}

The modified measure string action is

\begin{equation} \label{2}
S_{single-string} = -\int d \sigma d\tau \Phi [\frac12 \gamma^{ab}\partial_a X^{\mu} \partial_b X^{\nu} g_{\mu\nu}-  \frac{\epsilon^{ab}}{2\sqrt{-\gamma}}F_{ab}],
\end{equation}

where $F_{ab}$ is the field strength, $F_{ab} = \partial_a A_b - \partial_b A_a$, $A_a$ is the auxiliary Abelian gauge field. \\

This action is conformal invariant which is defined as

\begin{equation}
\varphi_a \rightarrow \varphi_a' = \varphi_a'(\varphi_a),
\end{equation}

\begin{equation}
\gamma_{ab} \rightarrow \Omega^2 \gamma_{ab},
\end{equation}

\begin{equation}
\Phi \rightarrow J \Phi = \Phi',
\end{equation}

such that $J = \Omega^2$, where $\Omega$ are the conformal transformations, $J$ is the Jacobian. \\

The constraint (\ref{eq:50}) implies

\begin{equation} \label{eq:1000}
\gamma^{cd} \partial_c X^{\mu} \partial_d X^{\nu} g_{\mu\nu} - \frac{\epsilon^{cd}}{\sqrt{-\gamma}} F_{cd}= M = const.
\end{equation}

If $M \ne 0$, we would obtain spontaneous breaking of conformal invariance but as we will see, $M$ turns out to be $0$. \\

The variations with respect to the dynamical fields $X^{\mu}, \gamma^{ab}, A_a$ of the action provide us with the equations of motion. \\

The field equations are

\begin{equation} \label{eq:7}
\partial_a(\Phi \gamma^{ab}\partial_bX^{\mu})+\Phi\gamma^{ab}\partial_a X^{\nu} \partial_b X^{\lambda} \Gamma^{\mu}_{\nu\lambda}=0,
\end{equation}

\begin{equation} \label{eq:06}
\partial_a X^{\mu} \partial_b X^{\nu} g_{\mu\nu} - \frac12 \gamma_{ab} \frac{\epsilon^{cd}}{\sqrt{-\gamma}} F_{cd}= 0,
\end{equation}

\begin{equation} \label{6}
\epsilon^{ab} \partial_b (\frac{\Phi}{\sqrt{-\gamma}}) = 0,
\end{equation}

Taking the trace of (\ref{eq:06}) and comparing with (\ref{eq:1000}) we get that $M = 0$ which means that the conformal invariance is not broken. By solving $\frac{\epsilon^{cd}}{\sqrt{-\gamma}}F_{cd}$ from (\ref{eq:1000}) (with $M = 0$) and introducing in (\ref{eq:06}), we obtain the sigma-model equations of motion, (\ref{eq:16}) and (\ref{eq:17}). The similarity of (\ref{eq:16}) and (\ref{eq:17}) derived from (\ref{1}) with (\ref{eq:7}) and (\ref{eq:06}) derived from (\ref{2}) proves the validity of the modified measure theory. \\

The tension is spontaneously induced. It is derived as a constant of integration from the equation (\ref{6}) \\

\begin{equation} \label{eq:201}
\frac{\Phi}{\sqrt{-\gamma}}= T.
\end{equation}

\section{The String Meson Model}

\begin{figure}[h]
\begin{center}
\includegraphics[width=0.4\textwidth,natwidth=410,natheight=242]{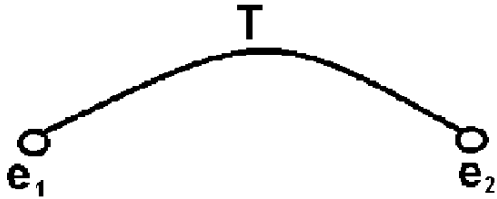}
\end{center}
\caption{The string meson. Circles are the charged endpoints. $T$ is the tension of the string. $e_1$ and $e_2$ are charges with the condition $e_1=-e_2$.}
\label{Pic100}
\end{figure}

A meson being a quark-antiquark bound system is reproduced by an open string with the opposite charged endpoints, see Fig.\ref{Pic100}. \\

This string is not infinite. A tension, $T$, is a constant along the string and vanishes at the endpoints. The string terminates when its tension discontinues, which follows from the following. \\

The modified measure string with the tension that can dynamically end at the endpoints has an action defined by

\begin{equation} \label{eq:6}
S_{single-endpoints} = S_{single-string} + \int d\sigma d\tau A_a j^a,
\end{equation}

where $j_a$ is the current of point-like charges setting at the endpoints. \\

The additional term in the action (\ref{eq:6}) contains the gauge field, $A_a$, interacting with point charges. Then the equations of motion with respect to the gauge field are modified compared to (\ref{6}). They are

\begin{equation} \label{eq:200}
\epsilon^{ab} \partial_b (\frac{\Phi}{\sqrt{-\gamma}}) = j^a.
\end{equation}

As the meson is in a static configuration then the current becomes

\begin{equation}
j^0 = \sum_i e_i \delta(\sigma - \sigma_i),
\end{equation}

where $e_i$ are charges that are associated with the gauge field $A^a$ and $\sigma_i$, $i=1,2$ are their locations, i.e. the endpoints. \\

Therefore,

\begin{equation}
\int d\sigma d\tau A_a j^a = \int d\sigma d\tau A_0(\tau, \sigma)j^0 = \sum_i e_i \int d \tau A_0(\tau, \sigma_i).
\end{equation}

Also (\ref{eq:200}) turns to

\begin{equation}
\epsilon^{01}\partial_1 (\frac{\Phi}{\sqrt{-\gamma}}) = j^0,
\end{equation}

where $(0,1)$ means $(\tau, \sigma)$. \\

Then, instead of (\ref{eq:201}) we obtain

\begin{equation} \label{eq:02}
\frac{\Phi}{\sqrt{-\gamma}} = \sum_i e_i \theta(\sigma - \sigma_i).
\end{equation}

For a proof we firstly consider the endpoint on the left with $e_1$ which is located at $\sigma_1$

\begin{equation}
\partial_{\sigma}(\frac{\Phi}{\sqrt{-\gamma}}) = e_1 \delta(\sigma - \sigma_1)
\end{equation}

and integrate the right-hand-side (rhs)

\begin{equation}
\int^{\sigma_1+\epsilon}_{\sigma_1-\epsilon} e_1 \delta(\sigma - \sigma_1)d\sigma = e_1,
\end{equation}

where $\epsilon$ is some positive constant. \\

The integration of the left-hand-side (lhs) gives

\begin{equation} \label{123}
\int^{\sigma_1+\epsilon}_{\sigma_1-\epsilon} \partial_{\sigma}(\frac{\Phi}{\sqrt{-\gamma}}) d\sigma = (\frac{\Phi}{\sqrt{-\gamma}})|_{\sigma_1+\epsilon} - (\frac{\Phi}{\sqrt{-\gamma}})|_{\sigma_1-\epsilon}.
\end{equation}

The first rhs term of (\ref{123}) is equal to $T$ because of (\ref{eq:201}) and the second one vanishes because the string starts at $\sigma_1$ and does not exist to $\sigma_1$'s left. Therefore, $e_1 = T$. \\

Now we consider the endpoint on the right with $e_2$ which is located at $\sigma_2$

\begin{equation}
\partial_{\sigma}(\frac{\Phi}{\sqrt{-\gamma}}) = e_2 \delta(\sigma - \sigma_2).
\end{equation}

In analogue with the endpoint on the left, we obtain

\begin{equation}
(\frac{\Phi}{\sqrt{-\gamma}})|_{\sigma_2+\epsilon} = e_2 + (\frac{\Phi}{\sqrt{-\gamma}})|_{\sigma_2-\epsilon} = 0.
\end{equation}

Then

\begin{equation}
T = -e_2.
\end{equation}

The tension is changed discontinuously from zero at one end of the string to some constant and back to zero at the other end of the string. This is the condition for the string to terminate. The opposite charges at the ends of the string guarantee it. \\

In the standard string theory the endpoints are free until the boundary conditions are applied. In the modified measure theory the boundary conditions are derived as is seen from the following consideration. \\

The variation with respect to $X^{\mu}$ gives the same equations of motion as those obtained from the action (\ref{2}). We need only the first term from the lhs of (\ref{eq:7}) because only this term could be singular. The external space is well defined and therefore so is $\Gamma^{\mu}_{\nu \alpha}$ and $\frac{\Phi}{\sqrt{-\gamma}}$ can jump but still remains finite but $\partial_a(\frac{\Phi}{\sqrt{-\gamma}})$ will be singular

\begin{equation} \label{eq:1}
\partial_a(\frac{\Phi}{\sqrt{-\gamma}})\sqrt{-\gamma}\gamma^{ab}\partial_bX^{\mu} = 0.
\end{equation}

Inserting (\ref{eq:02}) in (\ref{eq:1}) we obtain

\begin{equation}
e_i \delta(\sigma - \sigma_i)\delta^{\sigma}_a \sqrt{-\gamma} \gamma^{ab} \partial_b X^{\mu} = 0.
\end{equation}

The worldsheet metric $\gamma^{ab}$ can always be taken in a certain gauge to be conformally flat. Then in the conformal gauge in which $\gamma^{ab}\sqrt{-\gamma} = \eta^{ab}$, we obtain

\begin{equation} \label{12}
\partial_{\sigma} X^{\mu}(\tau, \sigma_i)=0.
\end{equation}

The equation (\ref{12}) is the Neumann boundary conditions, which are in fact the constraints on momentum components. They are obtained at the points where charges are located. Being originated from the discontinuity of the dynamical tension these conditions arise naturally in the framework of the modified measure theory. It is even impossible to violate them when having in hand only one string. \\







\section{The String Baryon Model}

\begin{figure}[h]
\begin{center}
\includegraphics[width=0.6\textwidth,natwidth=410,natheight=242]{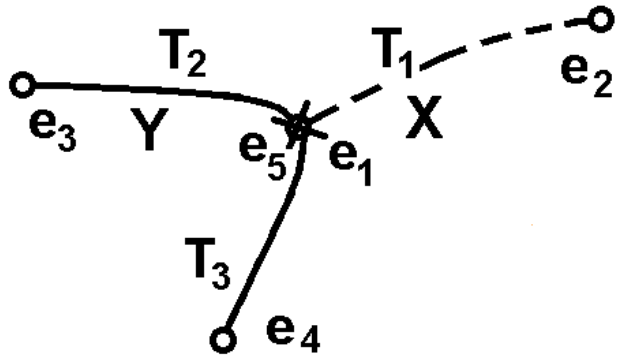}
\end{center}
\caption{The string baryon. Dotted and curved lines denote two strings, $X$ and $Y$, respectively. A cross is the intersection point. $T_1$, $T_2$, $T_3$ are the tensions. The sequence number of a charge is the same as the sequence number of its location, for example, the charge $e_3$ is located at the point $\sigma_3$.}
\label{Pic11}
\end{figure}

A baryon being a three-quark bound system is reproduced by two open strings with charged endpoints each and an additional charge within one of the strings, see Fig.\ref{Pic11}. \\

Note that the X-string stretches from $e_2$ to $e_1$ and the Y-string stretches from $e_3$ via $e_1$ to $e_4$. If one considers the line between $e_4$ (or $e_3$) via $e_1$ to $e_2$ as a Y-string and the line between $e_3$ (or $e_4$) and $e_1$ as an X-string, it will lead through the same calculations to the same results. Those choices are equivalent. However, once the choice is made, the picture is not symmetric anymore. We are working further with the one particular case because the charges are related among themselves and to the tensions in a particular way each time a choice is made. \\

The $X$-string has two endpoints with charges $e_1$ and $e_2$. In order for this string to terminate, the charges must be opposite: $e_1 = -e_2$. The $Y$-string too has two endpoints with charges $e_3$ and $e_4$. However, it has the additional charge $e_5$, which appears within the $Y$-string. Therefore, these strings do not enter equally. The charges $e_1$ and $e_5$ are located at the point $\sigma_1$ which is then the intersection point of two strings. At the point $\sigma_1$ the $Y$-string tension is changed from $T_2$ to $T_3$. Therefore, although $e_3 = T_2$ (see (\ref{eq:02})) but $T_3 = e_3 + e_5 = -e_4$. The mathematical formulation is coming. \\

Each string comes with its own internal metric $\gamma^{ab}_X$ or $\gamma^{ab}_Y$, its own measure $\Phi_X$ or $\Phi_Y$ and its own gauge field $A_a$ or $B_a$. \\

The additional terms in the $X$-string and $Y$-string actions are $\sum_i \int d\sigma d\tau A_a j^{ia}_A$ and $\sum_j\int d\sigma d\tau B_a j^{ja}_B$, respectively. Note that $i=1,2$ while $j=3,4,5$. The interaction takes place at $\sigma = \sigma_1$ as viewed from the $X$-string or $\sigma = \sigma_5$ as viewed from the $Y$-string. Then at this point ($\sigma = \sigma_1 = \sigma_5$) we obtain $j^0_A = e_1  \delta(\sigma - \sigma_1)$ and $j^0_B = e_5 \delta(\sigma - \sigma_1)$. \\

The $X$-string charges are similar to the string meson charges (\ref{12}) that were considered in the previous section. So are the $Y$-string endpoint charges. But the charge $e_5$ has its own Neumann boundary conditions at the point $\sigma_1$

\begin{equation}
\partial_{\sigma}(\frac{\Phi}{\sqrt{-\gamma}}) = e_5 \delta(\sigma - \sigma_1).
\end{equation}

Integrating both parts we obtain

\begin{equation}
\int^{\sigma_1+\epsilon}_{\sigma_1-\epsilon} e_5 \delta(\sigma - \sigma_1)d\sigma = e_5,
\end{equation}

\begin{equation}
\int^{\sigma_1+\epsilon}_{\sigma_1-\epsilon} \partial_{\sigma}(\frac{\Phi}{\sqrt{-\gamma}}) d\sigma = (\frac{\Phi}{\sqrt{-\gamma}})|_{\sigma_1+\epsilon} - (\frac{\Phi}{\sqrt{-\gamma}})|_{\sigma_1-\epsilon} = T_3 - T_2.
\end{equation}

Therefore

\begin{equation}
e_5 = T_3 - T_2.
\end{equation}

The equations of motion with respect to $Y^{\mu}$ gives us as previously the Neumann boundary conditions

\begin{equation}
\partial_{\sigma} Y^{\mu}(\tau, \sigma_1) = 0.
\end{equation}

All together the Neumann boundary conditions for the $Y$-string are

\begin{equation} \label{37}
\partial_{\sigma} Y^{\mu}(\tau, \sigma_j) = 0,
\end{equation}

where $j = 3,4,5$. \\

Then we see that the Neumann boundary conditions are applied not only at the endpoints but at the intersection point too. They signalize that the tension undergoes alterations: it becomes zero at the endpoints thereby terminating the string while at the intersection point it changes its value thereby dividing the string into two strings with different tensions. This is how we construct a baryon out of two strings as opposite to the more standard three string construction. \\

Just like the Neumann boundary conditions exist already in the modified measure theory, the Dirichlet boundary conditions are contained within the theory too and are derived in this section. \\

In order to obtain that $X$-string $=$ $Y$-string at $\sigma = \sigma_1$ the conditions for the intersection point are needed. \\

Our guiding principle is that the part of the action that is responsible for the interaction must be conformal invariant, generalizing the case of a single string equations. \\

The interaction term for two strings that leads to such conditions is

\begin{equation} \label{eq:101}
S_{interaction} = \int d\sigma d\tau(\lambda_1\sqrt{-\gamma_X} \gamma_X^{ab} + \lambda_2\sqrt{-\gamma_Y} \gamma_Y^{ab}) \partial_a(\frac{\Phi_X}{\sqrt{-\gamma_X}})\partial_b(\frac{\Phi_Y}{\sqrt{-\gamma_Y}}) V(X, Y),
\end{equation}

where $\lambda_1$, $\lambda_2$ are positive coefficients and $V(X,Y)$ is some potential that is defined later. The range of integration over $\sigma$ is taken to be $[-\infty, \infty]$, because, nonetheless, the physical range is only the parts where the tensions are not zero. The actual limits of integration are set dynamically. \\

Note that the $(\tau, \sigma)$-space being the common space of two strings is not the worldsheet of any string, and therefore, $\sigma$ is not the worldsheet coordinate. The actual space where the strings live is determined by measures, $\Phi_X$ and $\Phi_Y$. \\

The equations of motion provide us with the constraints on $V(X,Y)$. \\

Notice that in order to be effective $S_{interaction}$ requires both tensions, $\partial_a (\frac{\Phi_{X}}{\sqrt{-\gamma}_{X}})$ and $\partial_b (\frac{\Phi_{Y}}{\sqrt{-\gamma}_{Y}})$, to have a jump at the same point, otherwise $S_{interaction}$ vanishes. So that is why there is the need for the gauge fields charges at the point $\sigma_1$. \\

The equation (\ref{eq:101}) is indeed conformal invariant because

\begin{equation}
\gamma_X^{ab} \rightarrow \Omega^{-2} \gamma_X^{ab}, \quad \sqrt{-\gamma_X} \rightarrow \Omega^2 \sqrt{-\gamma_X}.
\end{equation}

Then

\begin{equation} \label{40}
\gamma_X^{ab} \sqrt{-\gamma_X} \rightarrow \gamma_X^{ab} \sqrt{-\gamma_X}.
\end{equation}

Using $\Phi_X \rightarrow \Omega^2 \Phi_X$, we obtain

\begin{equation}
\gamma_X^{ab} \Phi_X \rightarrow \gamma_X^{ab} \Phi_X
\end{equation}

and

\begin{equation}
(\frac{\Phi_X}{\sqrt{-\gamma_X}}) \rightarrow (\frac{\Phi_X}{\sqrt{-\gamma_X}}).
\end{equation}

Also

\begin{equation}
\gamma_Y^{ab} \sqrt{-\gamma_Y} \rightarrow \gamma_Y^{ab} \sqrt{-\gamma_Y}.
\end{equation}

Note that even through the full system has separate conformal invariance for each string, $\gamma^{ab}_{X} \rightarrow \Omega^{-2}_X \gamma^{ab}_X$, $\gamma^{ab}_{Y} \rightarrow \Omega^{-2}_Y \gamma^{ab}_Y$, the reparametrization invariance is still common. As we work in a common for both strings worldsheet spacetime, then a separate reparametrization invariance is not possible in principle . \\

The equations of motion with respect to $X^{\mu}$ acquire an extra term comparing to (\ref{eq:7}). It is

\begin{equation} \label{eq:102}
(\lambda_1 \sqrt{-\gamma}_{X} \gamma^{ab}_{X} + \lambda_2 \sqrt{-\gamma}_{Y} \gamma^{ab}_{Y}) \partial_a (\frac{\Phi_X}{\sqrt{-\gamma_X}}) \partial_b (\frac{\Phi_Y}{\sqrt{-\gamma_Y}}) \frac{\partial V(X, Y)}{\partial X^{\mu}}.
\end{equation}

From the analogues to (\ref{eq:02}) we see that

\begin{equation}
\partial_{\sigma} (\frac{\Phi_X}{\sqrt{-\gamma_X}}) = e_1 \delta(\sigma - \sigma_1),
\end{equation}

\begin{equation}
\partial_{\sigma} (\frac{\Phi_Y}{\sqrt{-\gamma_Y}}) = e_5 \delta(\sigma - \sigma_1).
\end{equation}

These terms produce two delta-functions, $\delta^2(\sigma - \sigma_1)$, which should be eliminated. Then

\begin{equation} \label{eq:100}
\frac{\partial V(X, Y)}{\partial X^{\mu}}|_{\sigma = \sigma_1} = 0.
\end{equation}

If not the X-string but Y-string is considered, then variation with respect to $Y^{\mu}$ gives us the similar condition

\begin{equation} \label{eq:10000}
\frac{\partial V(X, Y)}{\partial Y^{\mu}}|_{\sigma = \sigma_1} = 0.
\end{equation}

The equations of motion with respect to $\gamma_X^{cd}$ are altered too comparing with (\ref{eq:06}). The additional term is

\begin{equation} \nonumber
\lambda_1 \gamma_X^{ab} \frac12 \sqrt{-\gamma_X} \gamma_{Xcd} \partial_a (\frac{\Phi_X}{\sqrt{-\gamma_X}}) \partial_b (\frac{\Phi_Y}{\sqrt{-\gamma_Y}})V(X,Y)+
\end{equation}

\begin{equation} \nonumber
+ \lambda_1 \frac12 (\delta_a^c \delta_b^d + \delta_b^c \delta_a^d) \sqrt{-\gamma_X} \partial_a (\frac{\Phi_X}{\sqrt{-\gamma_X}}) \partial_b (\frac{\Phi_Y}{\sqrt{-\gamma_Y}})V(X,Y) +
\end{equation}

\begin{equation}
+ \frac12 \frac{\Phi_X}{\sqrt{-\gamma_X}}\gamma_{Xcd}\partial_a ((\lambda_1 \sqrt{-\gamma_X}\gamma_X^{ab} + \lambda_2 \sqrt{-\gamma_Y}\gamma_Y^{ab}) \partial_b (\frac{\Phi_Y}{\sqrt{-\gamma_Y}}) V(X,Y)) = 0.
\end{equation}

Since (\ref{eq:100}) is established, then the additional constraint on $V(X,Y)$ is

\begin{equation} \label{11111}
V(X, Y)|_{\sigma = \sigma_1} = 0.
\end{equation}

The intersection point is fixed now. The next task is to define the potential $V(X,Y)$ itself. \\














It is a function that is defined at the point where we demand the intersection of two strings to occur. The intersection condition is

\begin{equation} \label{eq:99}
X^{\mu}|_{\sigma = \sigma_1} = Y^{\mu}|_{\sigma = \sigma_1}.
\end{equation}

The most simple form it can take in the case of a flat spacetime background is

\begin{equation}
V = (X-Y)^2 = \eta_{\mu \nu}(X^{\mu} - Y^{\mu})(X^{\nu} - Y^{\nu}).
\end{equation}



The constraint (\ref{11111}) leads exactly to

\begin{equation}
X^{\mu}|_{\sigma = \sigma_1} = Y^{\mu}|_{\sigma = \sigma_1}.
\end{equation}

We obtain these conditions dynamically by adding the term to the action and specifying the potential $V(X,Y)$. These are exactly the Dirichlet boundary conditions. \\



\section{The String Baryon Model by 't Hooft \cite{h}}

Three strings ($X^{\mu,1}, X^{\mu,2}, X^{\mu,3}$) intersect at the point $\sigma = 0$. The Lagrange multipliers ($l_1^{\mu} (\tau), l_2^{\mu} (\tau)$) are introduced in the interaction term.

\begin{equation}
S_{interaction-tHooft} = \int d\tau (l_1^{\mu}(\tau)(X^{\mu, 1}(0, \tau)-X^{\mu, 3}(0, \tau)) + l_2^{\mu}(\tau)(X^{\mu, 2}(0, \tau)-X^{\mu, 3}(0, \tau))).
\end{equation}

The boundary conditions for the intersection point $\sigma = 0$ are

\begin{equation} \label{57}
\partial_{\sigma}(X^{\mu, 1} + X^{\mu, 2} + X^{\mu, 3}) = 0, \quad
X^{\mu, 1} - X^{\mu, 3} = X^{\mu, 2} - X^{\mu, 3} = 0,
\end{equation}

and for each endpoint ($\sigma = L^k(\tau)$) are:

\begin{equation} \label{59}
\partial_{\sigma} X^{\mu, k} = 0,
\end{equation}

where $k = 1,2,3$ and $L^k(\tau)$ are the lengths. \\

So the Neumann boundary conditions hold only for the sum $\sum_{k=1}^3 X^{\mu, k}$ at the intersection point. \\

By choosing conformal gauge, where $\gamma_{ab} = \Omega^2 \eta_{ab}$, the wave equation $\Box X^{\mu} = 0$ holds outside the intersection point or endpoints. Therefore

\begin{equation}
X^{\mu} = X_L^{\mu}(\tau + \sigma)+X_R^{\mu}(\tau - \sigma).
\end{equation}

Then the boundary conditions at the endpoints (\ref{59}) are

\begin{equation} \label{60}
X_L^{\mu, k}(\tau, L^k(\tau)) = X_R^{\mu, k}(\tau, L^k(\tau)).
\end{equation}

But in 't Hooft's treatment things are more complicated at the intersection point. The signal is propagated to the endpoints and reflects back to the intersection point. So that the boundary conditions (\ref{57}) are nonlocal in time and take the form

\begin{equation} \label{eq:001}
X_L^{\mu, k} (\tau, 0) = X_R^{\mu, k} (\tau_k, 0).
\end{equation}

As is seen, $X_L$ and $X_R$ are evaluated at different times, and $\tau_k(\tau)$ are the solutions of

\begin{equation}
\tau - \tau_k = 2L^k(\tau'_k), \quad
\tau'_k \equiv \frac{\tau + \tau_k}{2}.
\end{equation}

In the next section we show that this nonlocality is absent in our approach. \\

\section{The Resolution of the Nonlocality}

Starting from here $\sigma_0$ denotes the intersection point as previously interchangeably $\sigma_1$ and $\sigma_5$. We put $\sigma_0$ to $0$ for the comparison with 't Hooft's results. \\

The key feature of our model is that the Neumann boundary conditions (\ref{12}), (\ref{37}) hold not only at the endpoints but at the intersection point too. \\

Again by choosing the conformal gauge the wave equations $\Box X^{\mu} = 0$, $\Box Y^{\mu} = 0$ hold outside the intersection point or endpoints. Therefore

\begin{equation}
X^{\mu} = X_L^{\mu}(\tau + \sigma)+X_R^{\mu}(\tau - \sigma), \quad Y^{\mu} = Y_L^{\mu}(\tau + \sigma)+Y_R^{\mu}(\tau - \sigma).
\end{equation}

Then we directly obtain for the $X$-string:

\begin{equation}
X_L^{\mu, i}(\tau, 0) = X_R^{\mu, i}(\tau, 0)
\end{equation}

and for the $Y$-string:

\begin{equation}
Y_L^{\mu, j}(\tau, 0) = Y_R^{\mu, j}(\tau, 0).
\end{equation}

It is true up to a constant term that can be ignored while considered as either a function of $(\tau + \sigma)$ or $(\tau - \sigma)$ irrespectively. \\

As is seen, $X_L$ and $X_R$ are evaluated at the same time $\tau$. Therefore, we have locality at the intersection point as opposite to 't Hooft condition (\ref{eq:001}). \\

At the endpoints we still get the conditions (\ref{60}).

\section{The Solutions for the Equations of Motion in a Minkowski Background Spacetime}

We continue to assume that the endpoints are massless as opposed to \cite{y,x}, where the massive endpoints cases are investigated. Our analysis can be generalized for massive endpoints. \\

Here the rotation of the strings comes into play. We consider the motion on the plane, and two points are enough to define it. Any other motion demands higher dimensions and will unnecessary complicate our rotating configuration that is fully described in three dimensions. \\

As we are dealing with stringy particles, let's take the embedding spacetime to be the Minkowski spacetime. The signature of $\eta_{\mu\nu}$ is $(+1,-1,-1)$. \\

The equations of motion are

\begin{equation} \label{22222}
\Box X^{\mu} = \frac{1}{\sqrt{-\gamma}}\partial_{a}(\sqrt{-\gamma}\gamma^{ab}\partial_{b}X^{\mu})=0,
\end{equation}

where $\mu = 0,1,2$ denotes the components of $X$. As previously, all the calculations are correct for both branches of the $Y$-string too. \\

Note that since $T_1$, $T_2$, $T_3$ are not the same, then the wave vectors, $k_1$, $k_2$, $k_3$ ($k = \frac{2\pi}{\lambda}$), that will appear later, are not the same. \\

Variation of the sigma-model action with respect to $\gamma_{ab}$, the equation (\ref{eq:16}) , can be rewritten as

\begin{equation}
\gamma_{ab} = \frac{2\eta_{\mu\nu} \partial_a X^{\mu} \partial_b X^{\nu}}{\gamma^{cd}\partial_c X^{\mu}\partial_d X^{\nu} \eta_{\mu\nu}} = \Omega^2 h_{ab},
\end{equation}

where $h_{ab} = \eta_{\mu\nu}\partial_a X^{\mu} \partial_b X^{\nu}$ is the induced metric. As $\sqrt{-\gamma}\gamma^{ab}$ is invariant under conformal transformations (see equation (\ref{40})), then (\ref{22222}) reduces to

\begin{equation}
\Box X^{\mu} = \frac{1}{\sqrt{-h}}\partial_{a}(\sqrt{-h}h^{ab}\partial_{b}X^{\mu})= 0.
\end{equation}

As we will see there are solutions of the form

\begin{equation}
X^0 = c_1\tau + c_2 \sigma;
\end{equation}

\begin{equation}
X^1 = R(\sigma) \cos(\omega \tau);
\end{equation}

\begin{equation}
X^2 = R(\sigma) \sin(\omega \tau),
\end{equation}

where $c_1, c_2$ are some constants. \\

The Neumann boundary conditions are imposed at the intersection point ($\sigma = 0$). Therefore, $c_2 = 0$, and $X^0$ is a monotonic function of $\tau$:

\begin{equation}
X^0 = \tau.
\end{equation}

The Neumann boundary conditions are imposed at the endpoints too and provide that $\sigma = 0$. Then again

\begin{equation}
X^0 = \tau.
\end{equation}

Before the boundary conditions are imposed to $X^1$ and $X^2$, let's check that $R(\sigma)$ is an arbitrary function of $\sigma$. \\

The matrix elements are

\begin{equation} \nonumber
h_{\tau \tau} = \eta_{00} \partial_{\tau} X^0 \partial_{\tau} X^0 + \eta_{11} \partial_{\tau} X^1 \partial_{\tau} X^1 + \eta_{22} \partial_{\tau} X^2 \partial_{\tau} X^2=
\end{equation}

\begin{equation}
=1-(-R(\sigma) \sin(\omega \tau) \omega)^2  - (R(\sigma) \cos(\omega \tau) \omega)^2 = 1 - R^2(\sigma) \omega^2;
\end{equation}

\begin{equation} \nonumber
h_{\sigma \sigma} = \eta_{00} \partial_{\sigma} X^0 \partial_{\sigma} X^0 + \eta_{11} \partial_{\sigma} X^1 \partial_{\sigma} X^1 + \eta_{22} \partial_{\sigma} X^2 \partial_{\sigma} X^2 =
\end{equation}

\begin{equation}
= -(-(\frac{\partial R}{\partial \sigma})\sin(\omega \tau))^2 - ((\frac{\partial R}{\partial \sigma})\cos(\omega \tau))^2 = -(\frac{\partial R}{\partial \sigma})^2;
\end{equation}

\begin{equation}
h_{\tau \sigma} = h_{\sigma \tau} = 0.
\end{equation}

The inverse matrix elements are

\begin{equation}
h^{\tau\tau} = \frac{1}{1 - R^2(\sigma)\omega^2} ;
\end{equation}

\begin{equation}
h^{\sigma\sigma} = \frac{-1}{(\frac{\partial R}{\partial \sigma})^2};
\end{equation}

\begin{equation}
h^{\tau\sigma} = h^{\sigma \tau} = 0.
\end{equation}

Then

\begin{equation}
\det h_{\alpha \beta}= -(1 - R^2(\sigma) \omega^2)(\frac{\partial R}{\partial \sigma})^2.
\end{equation}

\begin{equation}
\sqrt{-\det h_{\alpha \beta}} \equiv \sqrt{-h} = \sqrt{(1 - R^2(\sigma) \omega^2)}(\frac{\partial R}{\partial \sigma}).
\end{equation}


Therefore, the equations of motion

\begin{equation}
\partial_{\tau} (\sqrt{-h} h^{\tau \tau} \partial_{\tau} X^{\mu}) + \partial_{\sigma} (\sqrt{-h} h^{\sigma \sigma} \partial_{\sigma} X^{\mu}) + \partial_{\tau} (\sqrt{-h} h^{\tau \sigma} \partial_{\sigma} X^{\mu}) + \partial_{\sigma} (\sqrt{-h} h^{\sigma \tau} \partial_{\tau} X^{\mu})= 0
\end{equation}

reduce to

\begin{equation} \label{eq:81}
\partial_{\tau} (\sqrt{-h} h^{\tau \tau} \partial_{\tau} X^{\mu}) + \partial_{\sigma} (\sqrt{-h} h^{\sigma \sigma} \partial_{\sigma} X^{\mu}) = 0.
\end{equation}

Then for $\mu = 0$:

\begin{equation}
\partial_{\tau}(\sqrt{1-R^2(\sigma)\omega^2}(\frac{\partial R}{\partial \sigma})\frac{1}{1-R^2(\sigma)\omega^2}) = 0
\end{equation}

for $\mu = 1$:

\begin{equation} \nonumber
\partial_{\tau}(\sqrt{1-R^2(\sigma)\omega^2}(\frac{\partial R}{\partial \sigma})\frac{1}{1-R^2(\sigma)\omega^2}(-R(\sigma)\omega \sin(\omega \tau))) +
\end{equation}

\begin{equation}
+\partial_{\sigma}(-\sqrt{1-R^2(\sigma)\omega^2}(\frac{\partial R}{\partial \sigma})\frac{1}{(\frac{\partial R}{\partial \sigma})^2}(\frac{\partial R}{\partial \sigma})\cos(\omega \tau)) = 0.
\end{equation}

for $\mu = 2$:

\begin{equation} \nonumber
\partial_{\tau}(\frac{\partial R}{\partial \sigma}\frac{1}{\sqrt{1-R^2(\sigma) \omega^2}}R(\sigma)\omega \cos(\omega \tau))+
\end{equation}

\begin{equation}
+ \partial_{\sigma}(\sqrt{1-R^2(\sigma) \omega^2}(\frac{\partial R}{\partial \sigma})\frac{-1}{(\frac{\partial R}{\partial \sigma})^2}(\frac{\partial R}{\partial \sigma})\sin(\omega\tau)) = 0.
\end{equation}

We do not specify $R(\sigma)$. However, the equations of motion (\ref{eq:81}) are satisfied. \\

As $R(\sigma)$ is arbitrary, we take it to be

\begin{equation} \label{1234567}
R(\sigma) = \sigma - \frac{1}{k} \sin(k\sigma).
\end{equation}

Next, the Neumann boundary conditions, $\frac{\partial X^{\mu}}{\partial \sigma} = 0$, are going to be set at the intersection point and then at the endpoints. \\

Again, at the intersection point $\sigma$ is assumed to be equal to $0$. Then \\

for $\mu = 1$ and $\mu = 2$:

\begin{equation}
X^1 = (\sigma - \frac{1}{k}\sin(k\sigma)) \cos({\omega\tau});
\end{equation}

\begin{equation}
X^2 = (\sigma - \frac{1}{k}\sin(k\sigma)) \sin(\omega \tau).
\end{equation}

At the endpoints $\sigma$ can not be assumed to be equal to $0$. Then \\

for $\mu = 1$ and $\mu = 2$:

\begin{equation}
X^1 = (\sigma - \frac{1}{k}\sin(k\sigma)) \cos({\omega\tau});
\end{equation}

\begin{equation}
X^2 = (\sigma - \frac{1}{k}\sin(k\sigma)) \sin(\omega \tau).
\end{equation}

Therefore, the Neumann boundary conditions at the endpoints give us the constraints on $k\sigma$ at the endpoints:

\begin{equation} \label{12345}
1 - \cos(k\sigma) = 0.
\end{equation}

Then the conditions for the wave vectors are

\begin{equation}
k \sigma = 2 \pi n,
\end{equation}

where $n$ is an integer. The intersection point is excluded, then $n \neq 0$. \\



The energy of a single string is

\begin{equation}
E = -p_0 = -\int P_0^{\tau} d\sigma,
\end{equation}

where $P_0^{\tau} \equiv \frac{\partial L}{\partial(\partial_{\tau} X^0)}$ and $L = -T \sqrt{(\partial_{\tau} X \partial_{\sigma} X)^2 - (\partial_{\tau} X)^2 (\partial_{\sigma} X)^2}$. Then in our case

\begin{equation} \nonumber
P_0^{\tau} = -T \frac{-(\partial_{\sigma} X)^2}{\sqrt{(\partial_{\tau} X \partial_{\sigma} X)^2 - (\partial_{\tau} X)^2 (\partial_{\sigma} X)^2}} =
\end{equation}


\begin{equation}
= -T (\frac{\partial R}{\partial \sigma})^2 \frac{1}{\sqrt{1-R^2(\sigma) \omega^2}(\frac{\partial R}{\partial \sigma})} = -T (\frac{\partial R}{\partial \sigma}) \frac{1}{\sqrt{1-R^2(\sigma)\omega^2}}.
\end{equation}

Therefore,

\begin{equation} \nonumber
E_{single} = T \int (\frac{\partial R}{\partial \sigma})\frac{1}{\sqrt{1-R^2(\sigma)\omega^2}}d\sigma = T \int_{R(\sigma_0=0)}^{R(\sigma_{endpoint})} \frac{d R}{\sqrt{1-R^2(\sigma)\omega^2}} =
\end{equation}

\begin{equation}
= - T \frac{1}{\omega} \arcsin(-\omega R(\sigma))|_{R(\sigma_0=0)}^{R(\sigma_{endpoint})} = T \frac{1}{\omega} \arcsin(\omega R(\sigma_{endpoint})).
\end{equation}

The energy of our string baryon configuration is

\begin{equation} \nonumber
E_{system} = T_1 \int_{R(\sigma_0=0)}^{R(\sigma_2)} \frac{d R}{\sqrt{1-R^2(\sigma)\omega^2}}+
\end{equation}

\begin{equation} \nonumber
+T_2 \int_{R(\sigma_0=0)}^{R(\sigma_3)} \frac{d R}{\sqrt{1-R^2(\sigma)\omega^2}}+ T_3 \int_{R(\sigma_0=0)}^{R(\sigma_4)} \frac{d R}{\sqrt{1-R^2(\sigma)\omega^2}}=
\end{equation}

\begin{equation}
= T_1 \frac{1}{\omega} \arcsin(\omega R(\sigma_2)) + T_2 \frac{1}{\omega} \arcsin(\omega R(\sigma_3)) + T_3 \frac{1}{\omega} \arcsin(\omega R(\sigma_4)).
\end{equation}

The angular momentum of a single string is

\begin{equation}
J^{\mu\nu}_{single} = -\int(X^{\mu} P^{\tau}_{\nu} - X^{\nu} P^{\tau}_{\mu}) d\sigma.
\end{equation}

In our case

\begin{equation}
J^{12}_{single} = -\int(X^1 P^{\tau}_2 - X^2 P^{\tau}_1) d\sigma,
\end{equation}

where

\begin{equation}
P_{1,2}^{\tau} = -T \frac{(\partial_\tau X \partial_{\sigma}X)\partial_{\sigma}X_{1,2} - (\partial_{\sigma} X)^2 \partial_{\tau} X_{1,2}}{\sqrt{(\partial_{\tau} X \partial_{\sigma} X)^2 - (\partial_{\tau} X)^2 (\partial_{\sigma} X)^2}}.
\end{equation}

Then

\begin{equation}
P_1^{\tau} = T (\frac{\partial R}{\partial \sigma}) \frac{R(\sigma) \omega \sin(\omega \tau)}{\sqrt{1 - R^2(\sigma) \omega^2}};
\end{equation}

\begin{equation}
P_2^{\tau} = -T (\frac{\partial R}{\partial \sigma}) \frac{R(\sigma) \omega \cos(\omega \tau)}{\sqrt{1 - R^2(\sigma) \omega^2}}.
\end{equation}

Therefore,

\begin{equation} \nonumber
J^{12}_{single} = -\int (-R(\sigma) \cos(\omega \tau) T (\frac{\partial R}{\partial \sigma}) \frac{R(\sigma)\omega \cos(\omega \tau)}{\sqrt{1-R^2(\sigma)\omega^2}} -
\end{equation}

\begin{equation} \nonumber
- R(\sigma)\sin(\omega \tau) T (\frac{\partial R}{\partial \sigma}) \frac{R(\sigma)\omega \sin(\omega \tau)}{\sqrt{1-R^2(\sigma)\omega^2}}) d \sigma = T \omega \int_{R(\sigma_0=0)}^{R(\sigma_{endpoint})} \frac{R^2(\sigma) d R}{\sqrt{1 - R^2(\sigma) \omega^2}} =
\end{equation}

\begin{equation} \nonumber
= T \omega(\frac{R(\sigma_{endpoint})}{-2 \omega^2}\sqrt{1 - R^2(\sigma_{endpoint}) \omega^2} + \frac{1}{2 \omega^2}(-\frac{1}{\omega} \arcsin(-\omega R(\sigma_{endpoint})))) =
\end{equation}

\begin{equation}
= T \frac{1}{2 \omega^2}(\arcsin(\omega R(\sigma_{endpoint})) -R(\sigma_{endpoint}) \sqrt{1 - R^2(\sigma_{endpoint}) \omega^2}).
\end{equation}

The angular momentum of our string baryon configuration is

\begin{equation} \nonumber
J^{12}_{system} = T_1 \frac{1}{2 \omega^2}(\arcsin(\omega R(\sigma_2)) -R(\sigma_2) \sqrt{1 - R^2(\sigma_2) \omega^2})+
\end{equation}

\begin{equation} \nonumber
+ T_2 \frac{1}{2 \omega^2}(\arcsin(\omega R(\sigma_3)) -R(\sigma_3) \sqrt{1 - R^2(\sigma_3) \omega^2}) +
\end{equation}

\begin{equation}
+ T_3 \frac{1}{2 \omega^2}(\arcsin(\omega R(\sigma_4)) -R(\sigma_4) \sqrt{1 - R^2(\sigma_4) \omega^2}).
\end{equation}




The Regge trajectory is a relation between the angular momentum and the square of the energy. Let's see it here. \\

Up to this point we did not assume any specific dynamics of the point charges at $\sigma_2, \sigma_3, \sigma_4$. Here we assign a massless dynamics to these endpoints. Therefore, $R(\sigma_2) \omega = R(\sigma_3) \omega = R(\sigma_4) \omega = 1$, that is, each endpoint moves with the speed of light. \\

From our choice of $R(\sigma)$, (\ref{1234567}), and the following condition on $k\sigma$, (\ref{12345}), we get

\begin{equation}
R(\sigma)|_{endpoint} = \sigma_{endpoint} = \frac{2 \pi n}{k}.
\end{equation}

Then for the lowest mode $(n=1)$, we obtain

\begin{equation}
R(\sigma)|_{endpoint} = \frac{1}{\omega}.
\end{equation}

Then

\begin{equation}
E_{system} = (T_1 + T_2 + T_3) \frac{1}{\omega} (\frac{\pi}{2});
\end{equation}

\begin{equation}
J^{12}_{system} = (T_1 + T_2 + T_3) \frac{1}{2 \omega^2}(\frac{\pi}{2}).
\end{equation}

Then the Regge trajectory of our system is

\begin{equation}
J =  2 \alpha' E^2,
\end{equation}

where $\alpha' = \frac{1}{2\pi (T_1 + T_2 + T_3)}$ is the slope parameter.

\section{Quantum Discussions}

In this paper we have ignored the structure of the quantum version of the theory. In this respect we propose this model as an effective model that may not be considered beyond the tree level, so the question of quantization may not be relevant. In any case any string model applied to hadron phenomenology has to be understood as an effective theory, since the fundamental theory is QCD, not a fundamental string theory. \\


It is interesting to notice that the structure of the action we have considered is linear in each measure. For example, the dependence on the measure $\Phi_X$ is (after integration by parts)

\begin{equation}
S = \int d\sigma d\tau \Phi_X L_X + \dots,
\end{equation}

With $L_X$ being independent of $\varphi_X^i$ ($\Phi_X = \epsilon^{ab}\epsilon_{ij}\partial_a \varphi^i_X \partial_b \varphi^j_X$), we get that the following infinite dimensional symmetry exists, up to a total divergence

\begin{equation}
\varphi_X^i \rightarrow \varphi_X^i + f^i(L_X),
\end{equation}

which preserves the linear structure of the action with respect to $\Phi_X$. This symmetry is infinite dimensional because it holds for any function $f^i(L_X)$, and the set of all functions is an infinite dimensional set. \\

Similar arguments can be made to justify the linearity of $\Phi_Y$. \\

Without any assumptions on $L$, we show now that $\varphi_i \rightarrow \varphi_i + f_i(L)$ is a symmetry. We do not specify whether $X$-string or $Y$-string is under consideration because the prof is similar for both strings. \\

\begin{equation} \nonumber
\Phi L = \epsilon^{ab}\epsilon_{ij} \partial_a \varphi_i \partial_b \varphi_j L \rightarrow \epsilon^{ab}\epsilon_{ij} \partial_a (\varphi_i + f_i(L)) \partial_b (\varphi_j + f_j(L)) L =
\end{equation}

\begin{equation} \nonumber
= \epsilon^{ab}\epsilon_{ij} \partial_a \varphi_i \partial_b \varphi_j L + \epsilon^{ab}\epsilon_{ij} \partial_a f_i(L) \partial_b \varphi_j L +
\end{equation}

\begin{equation}
+ \epsilon^{ab}\epsilon_{ij}\partial_a \varphi_i \partial_b f_j(L) L + \epsilon^{ab}\epsilon_{ij}\partial_a f_i(L) \partial_b f_j(L) L.
\end{equation}

Let's consider the last three terms separately: \\

The fourth term: \\

\begin{equation}
\epsilon^{ab}\epsilon_{ij}\partial_a f_i(L) \partial_b f_j(L) L = \epsilon^{ab}\epsilon_{ij}f_i' \partial_a L f_j' \partial_b L = 0,
\end{equation}

where $f_i'(L) = \frac{\partial f_i}{\partial L}$. It is equal to zero because $\partial_a L \partial_b L$ is symmetric, while $\epsilon^{ab}$ is antisymmetric. \\

The second and the third terms: \\

Let's define: \\

\begin{equation}
\epsilon^{ab}\epsilon_{ij} \partial_b \varphi_j = A_i^a,
\end{equation}

\begin{equation}
\epsilon^{ba}\epsilon_{ji} \partial_a \varphi_i = A_j^b.
\end{equation}

Then

\begin{equation} \nonumber
A_i^a f_i'(L) \partial_a L L + A_j^b f_j'(L) \partial_b L L = 2 A_i^a f_i'(L) \partial_a L L = 
\end{equation}

\begin{equation}
= \partial_a (A_i^a g_i(L)) = (\partial_a A^a_i) g_i(L) + A^a_i g_i'(L)\partial_a L,
\end{equation}

so since $\partial_a A^a_i = 0$, this is satisfied for

\begin{equation}
g_i'(L) = 2 f_i' (L) L,
\end{equation}

\begin{equation}
g_i(L) = 2\int dL f_i' (L) L.
\end{equation}

For every $f_i$ there exists such $g_i$. And the additional part of $L$ after the transformation is a total derivative. \\

This infinite dimensional symmetry (since the function $f_i(L)$ is arbitrary) is defined for all configurations, including those that do not satisfy equations of motion. It must be that way because in quantum theory we integrate over configurations that are off shell. \\

The existence of symmetries is usually used as a way to protect the theory, so that it keeps its basic structure even after quantum effects. Here the linearity on the measure, $\Phi$, in the action (so that it remains a measure) is protected under quantum corrections, in the case this symmetry (or a subgroup of this symmetry) is not plagued with anomalies. \\

Finally, other terms that do not contribute in the case of static configurations could be considered, like

\begin{equation}
\int d\sigma d\tau \epsilon^{ab} \partial_a (\frac{\Phi_X}{\sqrt{-\gamma_X}}) \partial_b (\frac{\Phi_Y}{\sqrt{-\gamma_Y}}) V(X,Y).
\end{equation}

This term could be relevant for quantum effects, like quantum creation processes, etc. This will be explored in a future publication.

\section{Conclusions}

We start our consideration with a single string. But instead of using $\sqrt{-\gamma}$ as a measure as it is done in a sigma-model action, we take a measure $\Phi$, which is constructed out of two scalar fields. We are permitted to do it as long as it is a density under arbitrary diffeomorphisms on the worldsheet spacetime, which is indeed the way we have constructed our measure. \\

Subsequently, the string tension appears as a constant of integration. In this framework it is not a scale that is put ad hoc but an additional dynamical degree of freedom. In \cite{x1} besides the supersymmetric extension, the gauge field and the new density in the action are quadratic and inverse, correspondingly, as opposed to the linear ones in our case. While our initial settings are different, the string tension appears there as an integration constant too. In principle the mechanism studied could be formulated also in the framework of the \cite{x1}'s approach, the action would be then modified by adding sources, etc. \\

We consider a single string. The charges at the endpoints of the string lead via the tension discontinuity to the Neumann boundary conditions. \\

Then we consider two strings. The endpoint of one string is connected to the internal part of the other one (See Fig.\ref{Pic11}). The charge in the internal part of the string lead via the tension alterations to the Neumann boundary conditions. By the addition of an interaction term to the modified action we obtain the conditions for the intersection that are the Dirichlet boundary conditions. \\

The action governing the string baryon configuration in Fig.\ref{Pic11} is

\begin{equation} \nonumber
S_{system} = -\int d\sigma d \tau \Phi_X [\frac12 \gamma_X^{ab}\partial_a X^{\mu} \partial_b X^{\nu} g_{\mu\nu}-  \frac{\epsilon^{ab}}{2\sqrt{-\gamma_X}}F_{ab}] + \sum_{i=1,2} \int d\sigma d\tau A_{a} j^{i}_A -
\end{equation}

\begin{equation} \nonumber
-\int d \tau d\sigma \Phi_Y [\frac12 \gamma_Y^{ab}\partial_a Y^{\mu} \partial_b Y^{\nu} g_{\mu\nu}-  \frac{\epsilon^{ab}}{2\sqrt{-\gamma_Y}}F_{ab}] + \sum_{j=3,4,5} \int d\sigma d\tau B_{a} j^{j}_B  +
\end{equation}

\begin{equation}
+ \int d\sigma d\tau(\lambda_1\sqrt{-\gamma_X} \gamma_X^{ab} + \lambda_2\sqrt{-\gamma_Y} \gamma_Y^{ab}) \partial_a(\frac{\Phi_X}{\sqrt{-\gamma_X}})\partial_b(\frac{\Phi_Y}{\sqrt{-\gamma_Y}}) V(X, Y),
\end{equation}

where

\begin{equation} \nonumber
F_{ab}^A = \partial_a A_b - \partial_b A_a,
\end{equation}

\begin{equation} \nonumber
F_{ab}^B = \partial_a B_b - \partial_b B_a.
\end{equation}

The $A$-gauge field couples directly to the measure of the $X$-string, and the $B$-gauge field couples directly to the measure of $Y$-string. The charges $e_1$ and $e_2$ belong to the $A$-field, the charges $e_3$, $e_4$ and $e_5$ belong to the $B$-field. The field strength $F_{ab}^A$ arises from the $A$ gauge field, and the field strength $F_{ab}^B$ arises from the $B$ gauge field. \\

Neumann boundary conditions are presented at all endpoints, $l = 1, 2$ $m = 3, 4$:

\begin{equation}
\partial_{\sigma} X^{\mu}(\tau, \sigma_l) = 0, \quad \partial_{\sigma} Y^{\mu}(\tau, \sigma_{m}) = 0.
\end{equation}

Both Dirichlet and Neumann boundary conditions are presented at the single intersection point.

\begin{equation}
X^{\mu}|_{\sigma = \sigma_5} = Y^{\mu}|_{\sigma = \sigma_5}, \quad \partial_{\sigma} Y^{\mu}(\tau, \sigma_5) = 0
\end{equation}

To avoid any confusion: generally, $\sigma$ denotes the location in the string. Especially, $\sigma_1$ and $\sigma_2$ denote the $X$-string endpoints with the charges $e_1$ and $e_2$, $\sigma_3$ and $\sigma_4$ denote the $Y$-string endpoints with the charges $e_3$ and $e_4$, $\sigma_5$ denotes the point in the $Y$-string where the charge $e_5$ is located. Strings intersect and by the construction, $\sigma_1$ and $\sigma_5$ denote the same location, the point of intersection. The charge $e_1$ being the endpoint of the $X$-string terminates the tension of the $X$-string and, as any other endpoint charge, raises the Neumann boundary conditions. The charge $e_5$ being the internal charge of the $Y$-string changes the value of the tension of the $Y$-string and raises the Neumann boundary conditions. Anytime, the tension of any string changes (including reducing to zero) the Neumann boundary conditions arise. In order for strings to intersect, the Dirichlet boundary conditions at the point $\sigma_1$ (the same as to say $\sigma_5$) are obtained. Starting from Section 6, where we make a comparison and later, when we solve the equations of motion, we denote the point of intersection as $\sigma_0$, that is $\sigma_1 = \sigma_5 = \sigma_0$ and without loss of generality set it to $0$. \\

The Neumann boundary conditions differ from the ones obtained in \cite{h}. There the Neumann boundary conditions at the intersection point hold only for the sum $\sum_{k=1}^3 X^{\mu, k}$. It leads to a nonlocality in (\ref{eq:001}). \\

A remarkable difference with \cite{h} is that in our case the boundary conditions at the intersection point become local in time:

\begin{equation}
X^{\mu, i}_R(\tau, 0) = X^{\mu, i}_L(\tau, 0)
\end{equation}

and

\begin{equation}
Y^{\mu, j}_R(\tau, 0) = Y^{\mu, j}_L(\tau, 0).
\end{equation}

Here the same $\tau$ is involved. This is due to the specific physics introduced to induce the boundary conditions: the dynamical tension mechanism which determines how strings end, interact etc. As we have seen, the approach with introduced Lagrange multipliers to enforce the strings to meet at some point $\sigma = 0$ is not equivalent. \\

The latter, as it is pointed out by 't Hooft himself, may require additional boundary conditions in particular because the method for propagating the signal from $\sigma = 0$ to $\sigma = L^k(\tau)$ and back can become ill-defined in some limits. Such problem is absent in our approach due to the locality in time of the boundary conditions. \\

In QCD which is the underlying microscopic theory in our case the chromo-electric field is generated by static quarks and leads to tube-like structures. String-like behavior in the chromo-electric field is well known. In our model charges too are responsible for the very existence of the string. It is demonstrated through the value of the string tension. In the string baryon model we use two types of abelian gauge fields. It is an indication that a more sophisticated theory that includes non-abelian gauge fields (and therefore many gauge fields automatically) would be more suitable. \\

Note that when the tensions are taken to be in a way that two of them are much greater than the third one then the diquark model arises. The introduction of the diquark provides us with a possibility to construct a more effective scheme for highly excited states, with less degrees of freedom and less number of highly excited states \cite{123}. \\

We have studied some rotating string solutions of equations of motion with the new boundary conditions. We obtain the energy and the angular momentum of our system, then assuming that each endpoint is a dynamical massless particle, the Regge trajectory is presented. A full analysis of the most general solutions will be done in the future. \\






\textbf{Acknowledgments}
TV acknowledges support by the Ministry of Aliyah and Integration (IL) and the Israel Science Foundation. EG is supported by the Foundational Questions Institute.

\end{document}